# Quantum Nonlinear Ferroic Optical Hall Effect

Hua Wang[1] and Xiaofeng Qian[1,*]

[1]Department of Materials Science and Engineering, Texas A&M University, College Station, Texas 77843, USA

*Correspondence to: feng@tamu.edu

**Abstract**:
Nonlinear optical responses to external electromagnetic field, characterized by second and higher order susceptibilities, play crucial roles in nonlinear optical devices and novel optoelectronics. Herein we present a quantum nonlinear ferroic optical Hall effect (QNFOHE) in multiferroics – a Hall-like direct photocurrent originated from the second order current response to intense electromagnetic field whose direction can be controlled by both internal ferroic orders and external light polarization. QNFOHE consists of two types of nonlinear photocurrent responses – shift current and circular photocurrent under linearly and circularly polarized light irradiation dominated by topological shift vector and Berry curvature, respectively. We elucidate the microscopic mechanism of QNFOHE in a representative class of 2D multiferroic materials using group theoretical analyses and first-principles electronic structure theory. The complex interplay of crystalline, permutation, gauge, and time reversal symmetries as well as inherent causality governs the symmetry properties of shift current and circular photocurrent. QNFOHE combined with rich ferroic degrees of freedom in multiferroic materials will open up a variety of new avenues for realizing tunable and configurable nonlinear optoelectronics etc.

**Introduction**
Nonlinear light-matter interaction plays a key role in the understanding, probing, and ultimate control of light and matter at the nanoscale. In particular, materials with strong nonlinear optical responses are highly desirable for many scientific disciplines and technological applications (*1-3*), e.g. ultrafast nonlinear optics (*4*), nonlinear biosensing and imaging (*5*), efficient generation of entangled photon pairs for quantum computing and quantum sensing (*6, 7*), and all-optical transistor and computer (*8-10*). Due to the odd parity of the two polar vectors – electric dipole and electric field, even order nonlinear electrical susceptibility tensors vanish in centrosymmetric crystals (*11, 12*), while odd order responses are not limited by such constraint.

Among a plethora of optical responses, nonequilibrium direct current (DC) from periodically driven system under light field is of particular interest. One notable example is linear photovoltaic Hall effect which has been predicted in graphene (*13*). Additionally, there exist appealing second-order nonlinear DC responses to electromagnetic field in noncentrosymmetric solids, e.g. shift current (SC) and circular photocurrent (CC). Both were recently observed in Weyl semimetals (*14-18*). In parallel, recent experiments suggest two-dimensional van der Waals layered materials exhibit symmetry-dependent strong nonlinear optical responses, such as second/third harmonic generation. It makes nonlinear optical spectroscopy a perfect facile tool for characterizing and quantifying 2D materials, e.g. elastic strain, crystal orientation, phase transition, crystal thickness, even/odd layer oscillation, etc. Strong nonlinear responses in 2D materials also suggest their great advantage for phase-matching free nonlinear optics (*19-21*).

SC, known as bulk photovoltaic current (*22, 23*), refers to a photoexcitation of an electron from the valence to the conduction band with a simultaneous change in the phase of wave functions. Another type of nonlinear photocurrent, CC, also known as injection current, circular photogalvanic effect (CPGE) (*24, 25*) and quantum nonlinear optical Hall current (*26, 27*), arises from the interference of wavefunctions upon photoexcitation associated with a phase difference between two linearly polarized light, which allows for phase-modulated nonlinear photocurrent with tunable magnitude and direction. For example, left-circularly polarized and right-circularly polarized light can induce opposite currents. Using semiclassical Boltzmann framework, Moore et al. (*26*) and Sodemann et al. (*27*) revealed the fundamental roots of nonlinear SC and CC in Berry curvature induced anomalous velocity of metallic materials. This framework includes intraband process as the product of Berry phase and the gradient of Fermi-Dirac distribution function, equivalent to Berry curvature dipole.(*27*) The nonlinear response in this case only considers the intraband process in metallic systems at low frequency regime. In intrinsic semiconductors or insulators, the gradient of Fermi-Dirac distribution and thus the Drude-like SC/CC response vanish. However, nonlinear photogalvanic current persists in non-centrosymmetric semiconductors due to nonlinear interband process which is absent in the above intraband model.

Herein we propose a *quantum nonlinear ferroic optical Hall effect* (QNFOHE) originating from the SC and CC responses in multiferroic materials, where a Hall-like direct photocurrent will be generated upon electromagnetic field whose direction can be controlled by both internal ferroic orders and external light polarization and chirality. We provide a microscopic picture of QNFOHE based on first-principles theory beyond semiclassical approach and group theoretical analysis of crystalline symmetry, time reversal symmetry, permutation symmetry, gauge symmetry, and inherent causality. To illustrate the underlying mechanisms, we take monolayer group IV monochalcogenides (MX with M=Ge, Sn and X=Se, S) as an example – a ferroelectric-ferroelastic multiferroics (*28*). These 2D MX exhibit large SC and CC responses that are dominated by topological quantities – shift vector and Berry curvature, respectively. Our first-principles calculations and group theory analyses show that switching ferroelastic order flips the direction of nonlinear SC and CC photocurrent by ±90º, while switching ferroelectric orders flips both nonlinear photocurrents by 180º. Additionally, changing left/right circular polarization of light will induce 180° flip in CC. The microscopic understanding of QNFOHE from first-principles theory, together with very recent discoveries of 2D ferroics/multiferroics, will open up a variety of new avenues for achieving tunable and configurable nonlinear optoelectronics etc. by utilizing their rich ferroic degrees of freedom.

**Results**

**Theory of SC and CC**
QNOFHE in 2D multiferroics originates from the second-order photo-induced direct current density $\langle \mathbf{J}_{DC}\rangle^{(2)}$. Unlike linear photocurrent, the direction of $\langle \mathbf{J}_{DC}\rangle^{(2)}$ depends on intrinsic ferroic orders which is the key to QNOFHE. $\langle \mathbf{J}_{DC}\rangle^{(2)}$ consists of two types of nonlinear photocurrents, SC and CC, which reflect the polarization change upon photoexcitation per unit volume,

$$\langle \mathbf{J}_{DC}\rangle^{(2)} = \langle \mathbf{J}_{SC}\rangle^{(2)} + \langle \mathbf{J}_{CC}\rangle^{(2)},$$

where

$$\langle J_{SC}^a\rangle^{(2)} = 2\sigma_2^{abc}(0;\omega,-\omega)E^b(\omega)E^c(-\omega),$$

$$\frac{d\langle J_{CC}^a\rangle^{(2)}}{dt} = -2\mathrm{Im}\,\eta_2^{abc}(0;\omega,-\omega)|E^b(\omega)||E^c(-\omega)|\sin(\phi^b - \phi^c).$$

Here $a,b,c$ are Cartesian indices. The electric field can be described using phasors $\mathbf{E}(t) = \mathbf{E}(\omega)\exp(-i\omega t) + c.c.$ For linearly polarized light $\mathbf{E}(\omega)$ is real, while for left/right-circularly polarized light $\mathbf{E}(\omega)$ is complex, $E^b(\omega) = |E^b(\omega)|e^{i\phi^b}$, with $\phi^b - \phi^c = \pm\frac{\pi}{2}$. We denote CC by $J_{CC}^{a,\circlearrowleft}$ and $J_{CC}^{a,\circlearrowright}$ for left and right-circularly polarized light, respectively, and denote SC by $J_{SC}^{a,\leftrightarrow}$ and $J_{SC}^{a,\updownarrow}$ for linearly x-/y-polarized light, respectively.

The SC susceptibility tensor $\sigma_2^{abc}$ can be derived from perturbation theory (*25*) as

$$\sigma_2^{abc}(0;\omega,-\omega) = \frac{i\pi e^3}{2\hbar^2}\int\frac{d\mathbf{k}}{8\pi^3}\sum_{nm\sigma}f_{nm}\left(r_{mn}^b(r_{nm}^c)_{;k^a} + r_{mn}^c(r_{nm}^b)_{;k^a}\right)\delta(\omega_{mn} - \omega),$$

where $(r_{nm}^b)_{;k^a} = \frac{\partial r_{nm}^b}{\partial k^a} - ir_{nm}^b(\mathcal{A}_n^a - \mathcal{A}_n^a)$ is the gauge covariant derivative. $\mathbf{r}_{nm} = i\langle n|\partial_\mathbf{k}|m\rangle$ and $\mathcal{A}_n = i\langle n|\partial_\mathbf{k}|n\rangle$ are interband and intraband Berry connection, respectively. $f$ is the Fermi-Dirac distribution with $f_{nm} \equiv f_n - f_m$, and $\Delta_{mn}^a \equiv v_{mm}^a - v_{nn}^a$ is the group velocity difference of band $m$ and $n$. The SC susceptibility tensor under linearly polarized light can be rewritten in a more elegant expression,

$$\sigma_2^{abb}(0;\omega,-\omega) = -\frac{\pi e^3}{2\hbar^2}\int\frac{d\mathbf{k}}{8\pi^3}\sum_{nm\sigma}f_{nm}R_{nm}^{a,b}(\mathbf{k})r_{nm}^b r_{mn}^b \delta(\omega_{mn} - \omega),$$

where $R_{nm}^{a,b}(\mathbf{k}) = -\frac{\partial \phi_{nm}^b(\mathbf{k})}{\partial k^a} + \mathcal{A}_n^a(\mathbf{k}) - \mathcal{A}_m^a(\mathbf{k})$ is shift vector and $\phi_{nm}(\mathbf{k})$ is the phase factor of the Berry connection $r_{nm}^b(\mathbf{k}) = |r_{nm}^b(\mathbf{k})|e^{i\phi_{nm}^b(\mathbf{k})}$. $r_{nm}^b r_{mn}^b$ is the optical absorption strength. Hence, SC is determined by the product of linear photoabsorption and topological shift vector integrated over the Brillouin zone.

The CC susceptibility tensor $\eta_2^{abc}$ (25) is given by
$$\eta_2^{abc}(0;\omega,-\omega) = -\frac{\pi e^3}{2\hbar^2} \int \frac{d\mathbf{k}}{8\pi^3} \sum_{nm\sigma} \Delta_{nm}^a f_{nm} [r_{mn}^b, r_{nm}^c] \delta(\omega_{mn} - \omega),$$
where $[r_{mn}^b, r_{nm}^c] \equiv r_{mn}^b r_{nm}^c - r_{mn}^c r_{nm}^b$, indicating CC vanishes under linearly polarized light. It can be rewritten in a general form assuming light propagates along z,
$$\eta_2^{a,z}(0;\omega,-\omega) = \frac{i\pi e^3}{2\hbar^2} \int \frac{d\mathbf{k}}{8\pi^3} \sum_{nm\sigma} f_{nm} \Delta_{nm}^a \Omega_{mn}^z(\mathbf{k}) \delta(\omega_{mn} - \omega).$$
$\Omega_{mn}^z(\mathbf{k}) \equiv i[r_{mn}^x, r_{nm}^y] = -i[r_{mn}^y, r_{nm}^x]$ is local Berry curvature between band $m$ and $n$. The global one reads $\Omega_m^z = \sum_{n \neq m} \Omega_{mn}^z$. It's clear to see $\eta_2^{abc} = -\eta_2^{acb}$. The original "$bc$" indices in the CC susceptibility tensor are now absorbed in the index "z".

SC and CC involve distinct physical processes. Figure 1 shows their microscopic picture using a two-band model. SC arises from the displacement of wave packet upon photoabsorption, while CC stems from the asymmetric transport of electrons and holes and the self-rotation of the wave packet. The latter induces orbital magnetic momentum coupled with the circularly polarized light. The intrinsic permutation symmetry of electric field leads to $\chi_{ijk}^{(2)}(-\omega_m - \omega_n; \omega_m, \omega_n) = \chi_{ikj}^{(2)}(-\omega_m - \omega_n; \omega_n, \omega_m)$. If time reversal symmetry is also present, $r_{nm}^b(-\mathbf{k}) = r_{mn}^b(\mathbf{k}), R_{nm}^{a,b}(-\mathbf{k}) = -R_{mn}^{a,b}(\mathbf{k}) = R_{nm}^{a,b}(\mathbf{k}), \mathbf{\Omega}_{mn}^z(\mathbf{k}) = -\mathbf{\Omega}_{mn}^z(-\mathbf{k})$. As a result of the causality and permutation symmetry, $\eta_2^{abc}(0;\omega,-\omega) = \eta_2^{acb}(0;-\omega,\omega) = -\eta_2^{abc}(0;-\omega,\omega) = [\eta_2^{abc}(0;-\omega,\omega)]^*$, ensuring $\eta_2^{abc}$ is purely imaginary. In contrast, $\sigma_2^{abc}(0;\omega,-\omega) = \sigma_2^{acb}(0;-\omega,\omega) = \sigma_2^{abc}(0;-\omega,\omega) = [\sigma_2^{abc}(0;\omega,-\omega)]^*$, suggesting $\sigma_2^{abc}$ is purely real.

**Relation between SC and CC**
Although SC and CC have different physical meaning, they are closely related. The derivative of CC susceptibility tensor $\eta_2^{a,z}$ can be written as the following
$$\partial_\omega \eta_2^{a,z}(0;\omega,-\omega) = \frac{i\pi e^3}{2\hbar^2} \int \frac{d\mathbf{k}}{8\pi^3} \sum_{nm\sigma} f_{nm} \Delta_{nm}^a \Omega_{mn}^z(\mathbf{k}) \partial_\omega \delta(\omega_{mn} - \omega)$$
$$= \frac{i\pi e^3}{2\hbar^2} \int \frac{d\mathbf{k}}{8\pi^3} \sum_{nm\sigma} f_{nm} (\partial_{k^a} \Omega_{mn}^z) \delta(\omega_{mn} - \omega),$$
where the integration by parts is applied. Furthermore,
$$\nabla_\mathbf{k} \times \mathbf{R}_{mn} \cdot \hat{\mathbf{z}} = \frac{\partial R_{mn}^{y,a}(\mathbf{k})}{\partial k^x} - \frac{\partial R_{mn}^{x,a}(\mathbf{k})}{\partial k^y} = \Omega_m^z - \Omega_n^z.$$

Thus for a two-band model, we substitute $\Omega_m^z - \Omega_n^z = 2\Omega_{mn}^z$
$$\partial_\omega \eta_2^{a,z}(0;\omega,-\omega) = \frac{-i\pi e^3}{4\hbar^2} \int \frac{d\mathbf{k}}{8\pi^3} \sum_\sigma \nabla_{k^a} (\nabla_\mathbf{k} \times \mathbf{R}_{vc} \cdot \hat{\mathbf{z}}) \delta(\omega_{cv} - \omega).$$

It clearly shows that the two topological quantities, shift vector and Berry curvature, are closely related. Both SC and the derivative of CC with respect to frequency are related to the shift vector.

**First-principles calculation and group theoretical analysis of nonlinear SC in 2D multiferroic MX**

The symmetry property of linear susceptibility and nonlinear SC and CC susceptibility are governed by point group and permutation symmetry, which correspond to direct product $\Gamma_P \otimes \Gamma_E$, $\Gamma_{J_{SC}} \otimes \Gamma_{EE}$, $\Gamma_{J_{CC}} \otimes \Gamma_{E \times E^*}$, respectively. Here we take monolayer group IV monochalcogenides (MX with M=Ge, Sn and X=Se, S) as an example which is a ferroelectric-ferroelastic multiferroics (28). In 2D MX with $C_{2v}$ point group (see Figs. 2A and 2B for crystal and electronic structure, respectively) and its character table (Table S1), we have
$$\Gamma_P \otimes \Gamma_E = (A_1 + B_1 + B_2) \otimes (A_1 + B_1 + B_2) = 3A_1 + 2A_2 + 2B_1 + 2B_2.$$
Hence there are three independent nonzero components in linear susceptibilities. The permutation symmetry further separates them into symmetric and asymmetric representations (29), $\Gamma_P \otimes \Gamma_E = \Gamma^s + \Gamma^a$, where $\Gamma^s = 3A_1 + A_2 + B_1 + B_2$ and $\Gamma^a = A_2 + B_1 + B_2$. Moreover, since polarization $P$, current $J$, and electric field $E$ are all polar vectors, they share the same representation, thus
$$\Gamma_{J_{SC}} \otimes \Gamma_{EE} = (A_1 + B_1 + B_2) \otimes (3A_1 + 2A_2 + 2B_1 + 2B_2) = 7A_1 + 6A_2 + 7B_1 + 7B_2,$$
$$\Gamma_{J_{SC}} \otimes \Gamma_{EE}^s = 5A_1 + 3A_2 + 5B_1 + 5B_2.$$
As a result, there are five independent nonzero components in the SC susceptibility tensor. Figures 2C and 2D show two of the five nontrivial frequency dependent SC susceptibilities $\sigma_2^{yxx}(0; \omega, -\omega)$ and $\sigma_2^{yyy}(0; \omega, -\omega)$ in 2D MX where linearly x- and y-polarized light are considered. The corresponding SC along y direction reads $J_{SC}^{y,\leftrightarrow} = 2\sigma_2^{yxx} E^x E^x$ and $J_{SC}^{y,\updownarrow} = 2\sigma_2^{yyy} E^y E^y$, respectively. The SC for monolayer GeS agrees well with the results in Refs. (21, 30). Furthermore, as shown in Fig. 2C and 2D, spin-orbit coupling (SOC) only slightly affects the SC in the case of GeS because of weak spin-orbit interaction strength of Ge and S atoms. $\sigma_2^{yyy}$ has two peaks below frequency of 3 eV. The first peak at 2 eV (denoted by red circle in Fig. 2C) is contributed from the k-points around the Brillouin zone center ($\Gamma$ point). The second peak at 2.8 eV denoted by green circle is apparently very strong. It comes from the transition at the k-points around the Brillouin zone boundary (X point) in a butterfly shape as shown in Fig. 2C. However, SOC can have significant impact in other cases such as well-known 1H-MoSe$_2$ and WSe$_2$, and the results for all MX and MX$_2$ without and with SOC are shown in Figs. S1 and S2 in the Supplementary Materials, respectively.

The frequency and reciprocal vector dependent contributions to SC are shown in Figs. 2E and 2F. They are determined by the product of SC susceptibility density, $\text{Im}[r_{vc}^b r_{cv;k^a}^b]$ (Figs. 2G and 2J), and the energy conservation law embedded in $\delta(\omega_{cv} - \omega)$. The distribution of shift current susceptibility density can be further understood by focusing on the frequency independent terms determined by the product of dipole transition strength, $r_{vc}^b r_{cv}^b$ (i.e. optical absorption), and shift vector $R_{nm}^{a,b}(\mathbf{k})$, as shown in Figs. 2H-2I and 2K-2L. $\text{Im}\, r_{vc}^x r_{cv}^x$ vanishes at the k points around the band gap, while $\text{Im}\, r_{vc}^y r_{cv}^y$ remains finite due to the optical selection rule. $R_{nm}^{a,b}(\mathbf{k})$ is a gauge invariant topological quantity which is well defined away from optical zero points, i.e. $r_{nm}^b(\mathbf{k}) \neq 0$. Since the optical zero points have no contribution to SC, we compute the shift vector by

$$R_{nm}^{a,b}(\mathbf{k}) = \frac{1}{\left|r_{nm}^{b}\right|^{2}} \text{Im}\left[r_{mn}^{b} r_{nm;k^{a}}^{b}\right].$$

The quantitative relationship between shift vector and polarization difference was recently reported. (31) In the presence of time reversal symmetry, $r_{nm}^{b}(-\mathbf{k}) = r_{mn}^{b}(\mathbf{k})$, $\left(r_{nm}^{b}(-\mathbf{k})\right)_{;k^{a}} = -\left(r_{mn}^{b}(\mathbf{k})\right)_{;k^{a}}$, which leads to $R_{nm}^{a,b}(-\mathbf{k}) = -R_{mn}^{a,b}(\mathbf{k}) = R_{nm}^{a,b}(\mathbf{k})$. This is clearly confirmed in Figs. 2I and 2L. The shift vector can reach as high as ~15 Å, much larger than its lattice constant. This is very different from electric polarization vector which is smaller than the lattice vector. Figures S3 and S4 show the distribution of the SC susceptibility tensor elements for monolayer MoSe$_2$ in the first Brillouin zone, demonstrating that its shift vector can go far beyond its lattice parameter.

**First-principles calculation and group theoretical analysis of nonlinear CC in 2D multiferroic MX**

For circularly polarized light along z, $(\mathbf{E} \times \mathbf{E}^{*})_{z}$ and axial axis $R_{z}$ share the $B_2$ representation. Therefore, $\Gamma_{j_x} \otimes \Gamma_{R_z} = B_2 \otimes B_2 = A_1$, indicating there is a nonzero CC response along x direction $\eta_2^{x,z}(0; \omega, -\omega)$ when the applied circularly polarized light is along z direction, *i.e.* perpendicular to 2D plane. Figures 3A and 3B show two antisymmetric CC susceptibility tensor elements, Im $\eta_2^{xxy}$ and Im $\eta_2^{xyx}$, respectively. Unlike SC, the main response of CC spread in two peaks from 2 to 6 eV. The peak values of Im $\eta_2^{abc}$ in the four MX materials are about $100 \sim 300 \times 10^8$ nm AV$^{-2}$s$^{-1}$, which allows us to estimate the generated nonlinear circular current under continuous wave limit as follows. At room temperature, a typical relaxation time of the electrons in MX materials is around $\tau = \mu_e / (\frac{e}{m^*}) \sim 10^{-14}s$. (32) $\mu_e$ is the mobility and $m^*$ is the effective mass of electrons. Considering a regular laser pointer with an intensity of 1mW/cm$^2$ and 2D MX with an effective thickness of 1nm, the induced circular photocurrent $J_{CC}$ can reach $10 \sim 30 \mu A/cm^2$, indicating the current can even be observed using low power continuous wave source (sheet photocurrent of $10 \sim 30$ nm $\frac{\mu A}{cm^2}$, that is, $100 \sim 300 pA/m$). (33) SC and CC are generated simultaneously under circularly polarized light. It is possible to compare their peak amplitudes if we assume the same incident light intensity and a relaxation time of $10^{-14}s$ for GeS. CC is larger than SC, $J_{CC}/J_{SC} = \sim 5$. The second order nonlinear photocurrent response for different incident polarized light is summarized in Table S3. It should be noted that there is another nonzero element $\eta_2^{z,x}(0; \omega, -\omega)$ in the CC susceptibility tensor since $\Gamma_{j_z} \otimes \Gamma_{R_x} = B_1 \otimes B_1 = A_1$, suggesting there is a CC response along z direction when the circularly polarized light is along x direction. However, $\eta_2^{z,x}$ is much smaller than $\eta_2^{x,z}$ discussed above.

Figures 3C and 3D show the $\mathbf{k}$ resolved CC susceptibility in monolayer GeS under circularly polarized light at two different frequencies (2.3 and 2.8 eV), demonstrating that the main response of the CC is localized around Y point. It should be noted that, for the same frequency of 2.8 eV, the SC (Fig. 2F) and CC (Fig. 3D) are very different from each other, as the SC is localized around Y point. The CC susceptibility is determined by the product of group velocity difference and Berry curvature $\mathbf{\Omega}_{cv}^{z}(\mathbf{k})$. The susceptibility tensor relates the component of the polar vector $\mathbf{J}$ and the axial vector $\mathbf{e} \times \mathbf{e}^{*}$. Hence, it is nonzero for the point groups that allow optical activity or gyrotropy. Here $\mathbf{e}$ is the unit vector of light polarization. Figure 3E shows the group velocity difference between the highest valence band and the lowest conduction band, which confirms the time reversal symmetry $\mathbf{\Delta}(\mathbf{k}) = -\mathbf{\Delta}(-\mathbf{k})$ (see Figs. S5 and S6 for the energy-dependent group velocity distribution). The Berry curvature $\mathbf{\Omega}_{cv}^z(\mathbf{k})$ of GeS is show in Fig. 3F, which confirms $\mathbf{\Omega}_{cv}^z(\mathbf{k}) = -\mathbf{\Omega}_{cv}^z(-\mathbf{k})$ under time reversal symmetry and the mirror plane (yz-plane) leads to opposite Berry curvature at $(\pm k^x, k^y)$. The product of these two odd functions, $\mathbf{\Delta}(\mathbf{k})$ and $\mathbf{\Omega}_{cv}^z(\mathbf{k})$, results in nonvanishing CC in 2D MX with $C_{2v}$ point group. This is in direct contrast to 1H-MoSe$_2$ whose CC response vanishes due to its D$_{3h}$ point group as evident in its Berry curvature shown in Fig. S7.

**Quantum nonlinear ferroic optical Hall effect (QNFOHE)**

The above group theoretical analyses and first-principles calculations illustrate the underlying selection rule and microscopic mechanism governing nonlinear SC and CC photocurrents. Since they are intimately related to the symmetry and topology, nonlinear SC and CC photocurrents are inherently coupled with the intrinsic ferroic orders in 2D multiferroics MX, giving rise to QNFOHE which we will discuss below. Let's first inspect the coupling between ferroelectric order ($P_y$) and nonlinear SC and CC responses. Since both CC and SC are a polar vector, the sign of SC will flip upon ferroelectric polarization switch ($P_y \to -P_y$). Consequently, the sign of SC susceptibility tensor $\sigma_2^{ybb}(0; \omega, -\omega)$ and CC susceptibility tensor $\eta_2^{x,z}(0; \omega, -\omega)$ will also flip. Thus, under the same linearly/circularly polarized light, SC and CC will change the direction by 180° upon ferroelectric polarization switch, that is, $J_{SC}^{y,\leftrightarrow}(P_y) = -J_{SC}^{y,\leftrightarrow}(-P_y)$, $J_{SC}^{y,\updownarrow}(P_y) = -J_{SC}^{y,\updownarrow}(-P_y)$, $J_{CC}^{x,\circlearrowleft}(P_y) = -J_{CC}^{x,\circlearrowleft}(-P_y)$, and $J_{CC}^{x,\circlearrowright}(P_y) = -J_{CC}^{x,\circlearrowright}(-P_y)$. Such property can be obtained from microscopic theory by considering the transformation rules of different matrix elements under space inversion and time reversal operation, including inter-band Berry connections $\mathbf{r}_{mn}(\mathbf{k})$, shift vector $\mathbf{R}_{mn}(\mathbf{k})$, group velocity difference $\mathbf{\Delta}_{mn}(\mathbf{k})$, and Berry curvature $\mathbf{\Omega}_{mn}(\mathbf{k})$ as listed in Table 1.

Now let's examine the coupling between ferroelastic order (e.g. spontaneous strain $\epsilon_{yy} > 0$ and $\epsilon_{xx} < 0$) and nonlinear SC and CC responses. Upon ferroelastic transition ($\epsilon_{yy} \to \epsilon_{xx}$ and $\epsilon_{xx} \to \epsilon_{yy}$), shift vector $R_{nm}^{y,b}(\mathbf{k})$, Berry curvature $\Omega_{mn}^z(\mathbf{k})$, $r_{nm}^b r_{mn}^b$ and $\Delta_{nm}^a$ will all rotate by 90°, which effectively switches the xy index. As a result, under the same linearly/circularly polarized light, nonlinear SC and CC photocurrent will change their direction by 90° upon ferroelastic transition, that is, $J_{SC}^{y,\leftrightarrow}(\epsilon_{yy}) = J_{SC}^{x,\updownarrow}(\epsilon_{xx})$, $J_{SC}^{y,\updownarrow}(\epsilon_{yy}) = J_{SC}^{x,\leftrightarrow}(\epsilon_{xx})$, $J_{CC}^{y,\circlearrowleft}(\epsilon_{yy}) = J_{CC}^{x,\circlearrowleft}(\epsilon_{xx})$, and $J_{CC}^{x,\circlearrowright}(\epsilon_{yy}) = J_{CC}^{x,\circlearrowright}(\epsilon_{xx})$. Note that in general $J_{CC}^{\circlearrowleft} = -J_{CC}^{\circlearrowright}$, $J_{SC}^{y,\leftrightarrow}$, and $J_{SC}^{y,\updownarrow}$ are independent. However, $J_{SC}^{y,\leftrightarrow} = -J_{SC}^{y,\updownarrow}$ holds in group D$_{3h}$ with a mirror plane perpendicular to $x$ axis, e.g. 1H-MoSe$_2$.

Since 2D MX possess both ferroelectric and ferroelastic orders, it has four multiferroic ($\pm P, \pm \epsilon$) states whose nonlinear photocurrent SC and CC are directly correlated as listed in Table 2. Here we define ferroelastic strain $+\epsilon$ for $\epsilon_{xx} < 0$ and $\epsilon_{yy} > 0$, and $-\epsilon$ for $\epsilon_{xx} > 0$ and $\epsilon_{yy} < 0$. Ferroelectric polarization $P$ could be $\pm P_x$ if $\epsilon_{xx} > 0$, or $\pm P_y$ if $\epsilon_{yy} > 0$. It is worth to mention that linear optical susceptibility will not change with ferroelectric polarization switching because its matrix element $r_{nm}^b r_{mn}^b$ is always positive. Moreover, for a given multiferroic state $(P, \epsilon)$, nonlinear SC and CC current responses in 2D MX are bulk photocurrent response along different directions, thus SC and CC can serve as a fundamental principle for real-space mapping of both ferroelectric and ferroelastic orders potentially at nanoscale resolution.

**Discussion**

In summary, using group theoretical analyses and first-principles calculations we have studied the microscopic mechanism of nonlinear SC and CC photocurrent in 2D multiferroics. Our results show that nonlinear photocurrent is highly sensitive to the symmetry of materials, including point group symmetry, permutation symmetry and time reversal symmetry. This leads to QNFOHE, a nonlinear optical Hall effect unique to multiferroics where the direct and sign of SC and CC photocurrent are strongly correlated with intrinsic ferroic orders and external light polarization. The concept of QNFOHE illustrated here is not limited to 2D multiferroics, rather it can be generally applicable to many multiferroic semiconductors and even ferroelectric metals. With QNFOHE one can envisage to directly control nonlinear photocurrent by switching ferroelastic strain and/or ferroelectric polarization accompanied by instantaneous direction and/or sign switching of the photocurrents. One may conduct high-resolution characterization of ferroelastic and ferroelectric orders as well as domain evolution using ultrafast optical techniques based on QNFOHE. The QNFOHE together with the recently discovered 2D ferroics/multiferroics will open up avenues for realizing configurable nonlinear optoelectronics etc. by controlling their rich ferroic orders.

**Acknowledgments:**

**Funding:** This research was supported by NSF under award number DMR-1753054. Portions of this research were conducted with the advanced computing resources provided by Texas A&M High Performance Research Computing.

**Author contributions:** X.Q. conceived the project. H.W. and X.Q. developed first-principles codes for computing second order nonlinear shift and circular photocurrent. H.W. performed DFT simulations. H.W. and X.Q. analyzed the results and wrote the manuscript.

**Competing interests:** The authors declare that they have no competing interests.

**Figures and Tables**

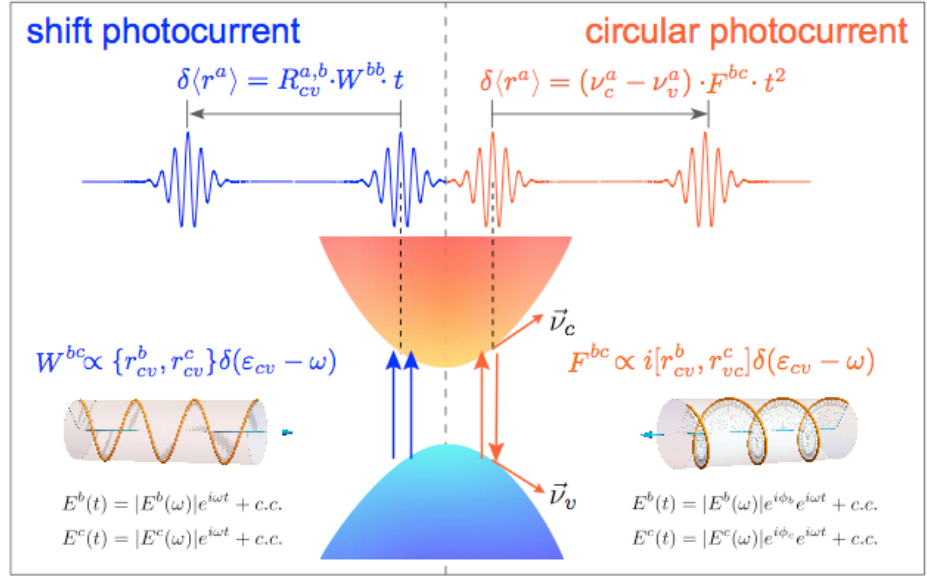

**Fig. 1. Microscopic interpretation of shift current and circular current using two-band model.** $\delta\langle r^a \rangle$ is the variation of the mean value of position operator indicating the shift of electron wave packet in real space. $W$ is the linear optical transition rate at frequency $\omega$. Photoexcitation induces shift of the center of the electron wave packet in real space. SC comes from the displacement of wave packet upon photoabsorption, while CC stems from the asymmetric motion of electrons and holes and the self-rotation of the wave packet. The latter induces itinerant orbital magnetic momentum coupled with the circularly polarized light.

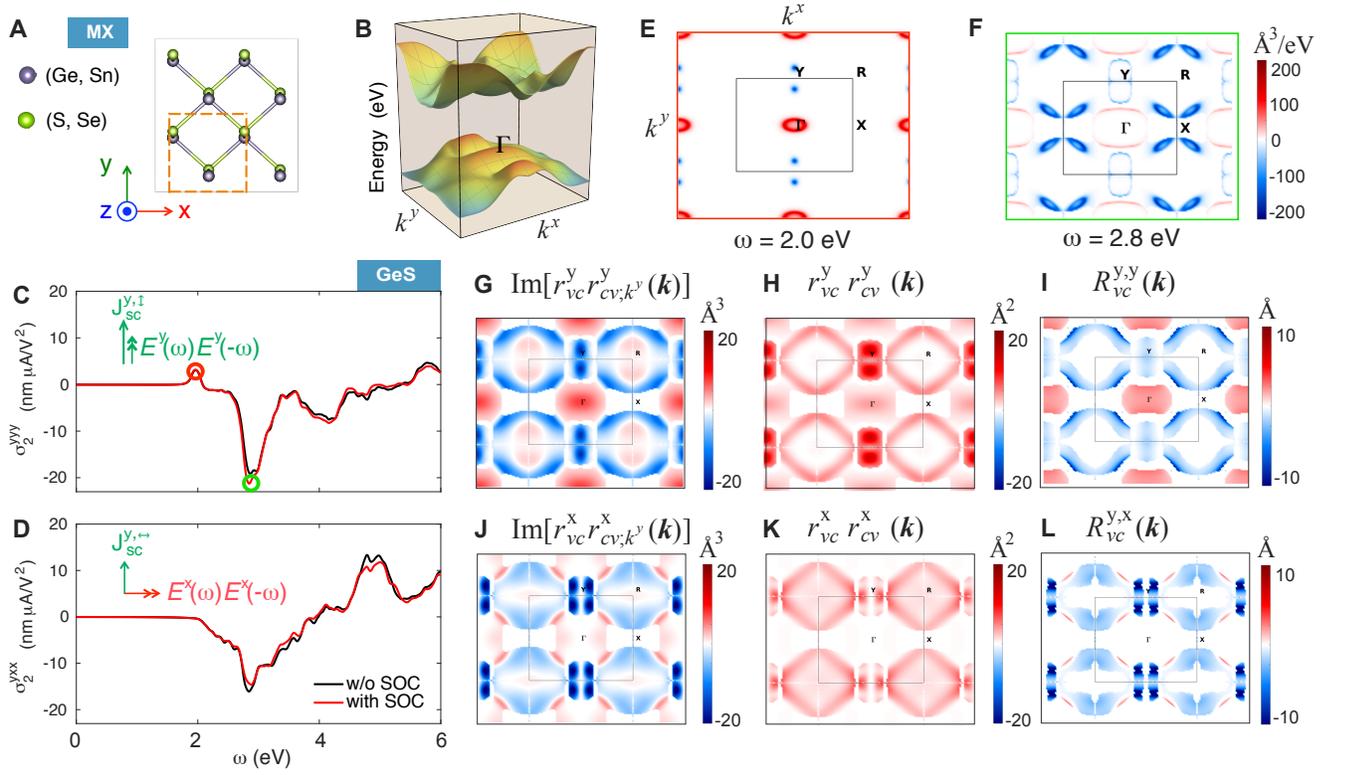

**Fig. 2. Shift current (SC) and its microscopic origin in 2D ferroelastic-ferroelectric monolayer group IV monochalcogenide GeS.** (**A**) Crystal structure of monolayer group IV monochalcogenides MX where M=(Ge, Sn) and X=(S, Se). (**B**) 2D electronic band structure near the Fermi level. (**C**) and (**D**) Frequency-dependent nonlinear shift current response to incoming linearly x and y polarized light, respectively. (**E**) and (**F**) Reciprocal vector ***k*** resolved SC susceptibility under linearly y polarized light at the first two peaks (2.0 and 2.8 eV). (**G**) and (**J**) ***k***-resolved SC strength, (**H**) and (**K**) ***k***-resolved dipole transition strength, (**I**) and (**L**) ***k***-resolved topological shift vector of 2D GeS in 2D Brillouin zone under linearly x and y polarized light, respectively.

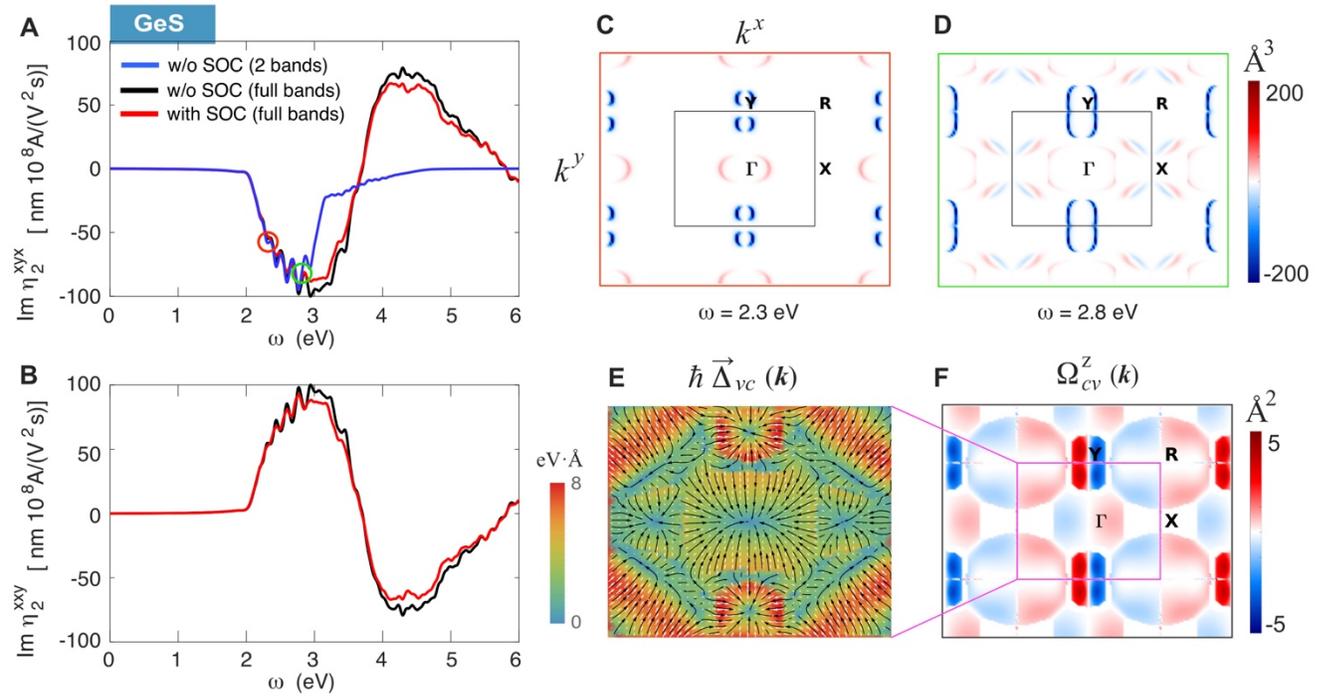

**Fig. 3. Circular photocurrent (CC) and its microscopic origin in 2D ferroelastic-ferroelectric monolayer group IV monochalcogenide GeS.** (**A**) and (**B**) Two opposite CC susceptibility tensor elements induced by circularly-polarized light. (**C**) and (**D**) Evolution of reciprocal vector ***k*** resolved CC susceptibility under circularly polarized light at two different frequencies (2.3 and 2.8 eV). (**E**) and (**F**) Group velocity difference and Berry curvature between the highest valence band and the lowest conduction band. The white arrows in (E) denote the calculated group velocity different at specific ***k*** point, and the black curves indicate the associated stream lines.

| Quantity | Symmetry Operation | |
|---|---|---|
| | Space Inversion, $\mathcal{J}$ | Time Reversal, $\mathcal{T}$ |
| $\mathbf{r}_{mn}(\mathbf{k})$ | $-\mathbf{r}_{mn}(-\mathbf{k})$ | $\mathbf{r}^*_{mn}(-\mathbf{k})$ |
| $\mathbf{R}_{mn}(\mathbf{k})$ | $-\mathbf{R}_{mn}(-\mathbf{k})$ | $\mathbf{R}_{mn}(-\mathbf{k})$ |
| $\mathbf{\Delta}_{mn}(\mathbf{k})$ | $-\mathbf{\Delta}_{mn}(-\mathbf{k})$ | $-\mathbf{\Delta}_{mn}(-\mathbf{k})$ |
| $\mathbf{\Omega}_{mn}(\mathbf{k})$ | $\mathbf{\Omega}_{mn}(-\mathbf{k})$ | $-\mathbf{\Omega}_{mn}(-\mathbf{k})$ |

**Table 1. Transformation of inter-band Berry connections $\mathbf{r}_{mn}$, shift vector $\mathbf{R}_{mn}$, group velocity difference $\mathbf{\Delta}_{mn}$ and Berry curvature $\mathbf{\Omega}_{mn}$ under space inversion $\mathcal{J}$ and time reversal $\mathcal{T}$ symmetry operation.** $\mathbf{R}_{mn}(\mathbf{k})$ is odd under $\mathcal{J}$ and even under $\mathcal{T}$ in moment space. $\mathbf{\Omega}_{mn}(\mathbf{k})$ is even under $\mathcal{J}$ and odd under $\mathcal{T}$ in moment space. These transformation rules govern the coupling between ferroelectric polarization and nonlinear SC and CC photocurrent: $J^{y,\leftrightarrow}_{SC}(P_y) = -J^{y,\leftrightarrow}_{SC}(-P_y), J^{y,\updownarrow}_{SC}(P_y) = -J^{y,\updownarrow}_{SC}(-P_y), J^{x,\circlearrowleft}_{CC}(P_y) = -J^{x,\circlearrowleft}_{CC}(-P_y)$, and $J^{x,\circlearrowright}_{CC}(P_y) = -J^{x,\circlearrowright}_{CC}(-P_y)$.

| Nonlinear optical Hall current $(J_x, J_y)$ | | | ferroelectric order | | | |
|---|---|---|---|---|---|---|
| | | | +P | | −P | |
| | | | $+P_x$ if $\epsilon_{xx} > 0$ | $+P_y$ if $\epsilon_{yy} > 0$ | $-P_x$ if $\epsilon_{xx} > 0$ | $-P_y$ if $\epsilon_{yy} > 0$ |
| ferroelastic order | $+\epsilon$ | $\epsilon_{xx} < 0$ & $\epsilon_{yy} > 0$ | $(J^{\circlearrowleft,\circlearrowright}_{CC}, J^{\leftrightarrow,\updownarrow}_{SC})$ | | $(-J^{\circlearrowleft,\circlearrowright}_{CC}, -J^{\leftrightarrow,\updownarrow}_{SC})$ | |
| | $-\epsilon$ | $\epsilon_{xx} > 0$ & $\epsilon_{yy} < 0$ | $(J^{\updownarrow,\leftrightarrow}_{SC}, J^{\circlearrowleft,\circlearrowright}_{CC})$ | | $(-J^{\updownarrow,\leftrightarrow}_{SC}, -J^{\circlearrowleft,\circlearrowright}_{CC})$ | |

**Table 2. Quantum nonlinear ferroic optical Hall effect (QNFOHE).** Second-order nonlinear photocurrent SC and CC responses are directly correlated with the intrinsic ferroic orders $(\pm P, \pm \epsilon)$ of 2D MX materials and external linear ($\leftrightarrow, \updownarrow$) and circular ($\circlearrowleft, \circlearrowright$) polarization of incoming light. Total 16 types of in-plane nonlinear photocurrents can be generated by controlling four ferroic states and four types of light polarizations.